\documentclass[12pt]{article}
\usepackage{graphicx}
\usepackage{color}
\newcommand\pubnumber{NuPhys2018-Gando}
\newcommand\pubdate{\today}

\def\napoli{Research Center for Neutrino Science, Tohoku University\\
Sendai 980-8578, JAPAN}
\def\Title#1{\begin{center} {\Large #1 } \end{center}}
\def\Author#1{\begin{center}{ \sc #1} \end{center}}
\def\Address#1{\begin{center}{ \it #1} \end{center}}

\newcommand\pubblock{\rightline{\begin{tabular}{l} \pubnumber\\
         \pubdate  \end{tabular}}}
\newenvironment{Abstract}{\begin{quotation}  }{\end{quotation}}
\newenvironment{Presented}{\begin{quotation} \begin{center} 
             PRESENTED AT\end{center}\bigskip 
      \begin{center}\begin{large}}{\end{large}\end{center} \end{quotation}}
\def\Acknowledgements{\bigskip  \bigskip \begin{center} \begin{large}
             \bf ACKNOWLEDGEMENTS \end{large}\end{center}}
\def\beq{\begin{equation}}
\def\eeq#1{\label{#1}\end{equation}}
\def\eeqn{\end{equation}}

\def\beqa{\begin{eqnarray}}
\def\eeqa#1{\label{#1}\end{eqnarray}}
\def\eeqan{\end{eqnarray}}

\let\bar=\overbar

\def\Dslash{\not{\hbox{\kern-4pt $D$}}}
\def\dslash{\not{\hbox{\kern-2pt $\del$}}}

\def\msb{{\bar{\ssstyle M \kern -1pt S}}}

\begin{document}
\begin{titlepage}
\pubblock

\vfill
\Title{Neutrinoless double beta decay search with liquid scintillator experiments}
\vfill
\Author{Yoshihito Gando\\ for the KamLAND-Zen Collaboration}
\Address{\napoli}
\vfill
\begin{Abstract}
Liquid scintillator experiments for neutrinoless double beta decay search have high sensitivity 
based on the ultra low background environment and high scalability.
This paper describes an overview of current ongoing projects KamLAND-Zen and SNO+.
\end{Abstract}
\vfill
\begin{Presented}

NuPhys2018, Prospects in Neutrino Physics

Cavendish Conference Centre, London, UK, December 19--21, 2018

\end{Presented}
\vfill
\end{titlepage}
\def\thefootnote{\fnsymbol{footnote}}
\setcounter{footnote}{0}

\section{Introduction}

Neutrinoless double beta ($0\nu\beta\beta$) decay is a key for physics beyond the Standard Model of elementary particles. 
If we observe this decay, then the neutrino behaves as a Majorana particle\cite{Majorana} and the decay process 
violates lepton number conservation. 
If the neutrino is Majorana type, extremely light neutrino mass is explained via the seesaw mechanism\cite{seesaw}, 
and may also explain the baryon asymmetry via leptogenesis\cite{leptogenesis}.
The decay rate of $0\nu\beta\beta$ ($\left( T_{1/2}^{0\nu} \right)^{-1}$) is proportional to the square of the effective neutrino mass $\langle m_{\beta\beta} \rangle$ 
as follows,
\begin{eqnarray}
    \left( T_{1/2}^{0\nu} \right)^{-1}
    = G^{0\nu} \left| M^{0\nu} \right|^2 \langle m_{\beta\beta} \rangle^2
\end{eqnarray}  
where $ \left| \langle m_{\beta\beta} \rangle \right| \equiv \left|  \left| U_{e1}^{L}\right|^2 m_1 + \left| U_{e2}^{L}\right|^2 m_2 e^{i\phi_2} +  \left| U_{e3}^{L}\right|^2 m_3 e^{i\phi_3} \right|$,
$T_{1/2}^{0\nu}$ is the half-life,  $G^{0\nu}$ is the phase space factor, $M^{0\nu}$ is the nuclear matrix element, $e^{i\phi_{2,3}}$ are Majorana CP phases, and $U^L_{ej}$ ($j$ = 1 - 3) is the neutrino mixing matrix. 
The event rate determines the mass scale of light neutrino mass.

$0\nu\beta\beta$ emits two beta rays and the total energy corresponds to the Q-value of the double beta decaying nucleus.
Usually the Q-values are in the region of environmental backgrounds caused by uranium-chain and thorium decay chains. 
The sensitivity to $\langle m_{\beta\beta} \rangle$ is proportional to the square root of the exposure time for a background-free case, 
but it will be reduced to the fourth root of the exposure time in background limited cases\cite{2to4square}.
Therefore, in order to observe the $0\nu\beta\beta$ signal, we need many double beta decaying nuclei, a long live time, and a background-free environment or powerful background rejection methods to eliminate noise events.

Recently there are four types of experiments used in $0\nu\beta\beta$ searches. 
The first type are high energy resolution detector ($\sim$0.1\%) using germanium detectors~\cite{GERDA, MAJORANA} or bolometers~\cite{CUORE}.
These detectors can reduce backgrounds in the observed energy spectrum. 
The second type are tracking detectors. 
There the source and the detector are separated, thus it is hard to contain a large amount of nuclei, 
but the event pattern of 2$\beta$ decay can be identified~\cite{NEMO}. 
The third type are xenon TPC detectors~\cite{EXO,NEXT,PandaX,AXEL}. 
This type has both previous features partially, good energy resolution ($\sim$0.1\%; for gas, $\sim$3\%; for liquid) 
and high event pattern identification by TPC. 
The final type are liquid scintillator detectors~\cite{Zen400final}\cite{SNO+}.
These detectors have poor energy resolution ($\sim$10\%) with no particle identification methods for $\beta/\gamma$.
However, liquid scintillator detectors realize ultra low background environments for radiation from uranium, thorium, 
and other metals which are used in the detector or the vessel, 
and can contain a large amount of double beta decaying nuclei.
Thus this type is one of the most sensitive detectors for $0\nu\beta\beta$ search.
In this paper, we report the current status of liquid scintillator (LS) detector experiments, KamLAND-Zen and SNO+.

\section{Liquid scintillator detector}
\begin{figure}[htb]
\centering
\includegraphics[height=2.0in]{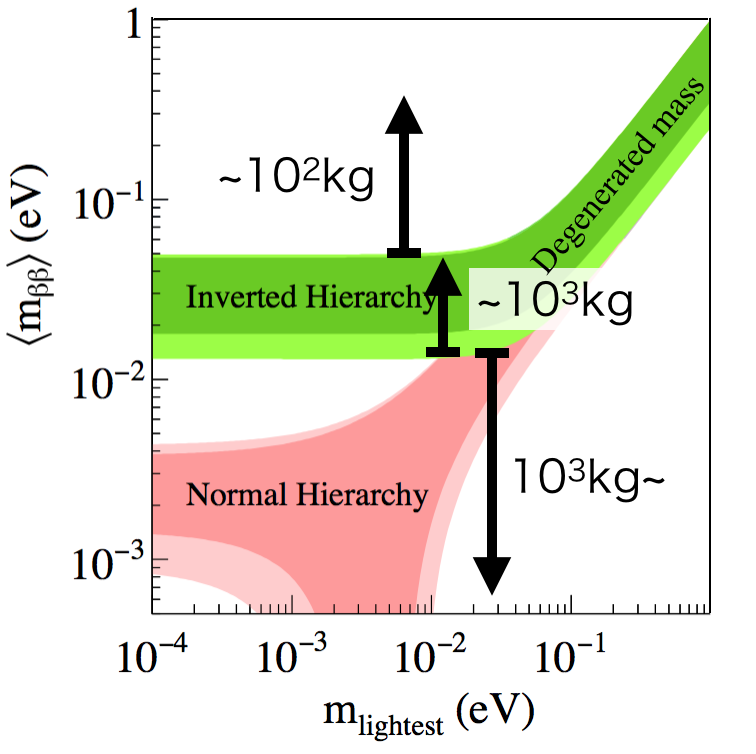}
\vspace{-5mm}
\caption{Allowed region of effective neutrino mass as a function of the lightest neutrino mass.}
\label{fig:alowedregion}
\end{figure}
Liquid scintillator experiments for $0\nu\beta\beta$ decay were originally developed for neutrino oscillation 
experiments at MeV energy region. 
KamLAND-Zen and SNO+ have similar designs.
Inward looking photomultiplier tubes (PMTs) are set on the inner surface of $\sim$18m diameter stainless tank 
or structure and $\sim$1,000 tons of LS is stored in a 13 m diameter nylon/EVOH base balloon (KamLAND-Zen) 
or 12 m diameter acrylic sphere (SNO+). 
The emitted scintillation light caused by radiation in LS ($\alpha, \beta, \gamma$, etc.) are detected by PMTs. 
The vertex is reconstructed by hit timing and the energy is reconstructed by transparency corrected charge of PMTs. 
Liquid scintillator is purified by water extraction, distillation, nitrogen purge etc., and the contamination level for uranium and 
thorium can reach a sufficient level of $O(10^{-18})$ g/g.
Thus the cleanness of the container for $\beta\beta$ decaying nuclei loaded liquid scintillator or a large 
self shielding distance from the surface of the container are the key elements for a high sensitivity search for $0\nu\beta\beta$ decay.

Other possible backgrounds in liquid scintillator detectors are spallation products of carbon caused 
by cosmic ray muons, solar $^8$B neutrinos, and energy tail of $2\nu\beta\beta$ decay spectrum.
If the detector site is very deep and the muon event rate is low, $^{10}$C background made by spallation is negligible.
However if this rate is high, it has to be rejected, for example, by triple coincidence between muon, 2.2 MeV $\gamma$-ray from neutron capture by proton, and $^{10}$C decay ($\beta^{+}$, $\tau$ = 27.8 s, Q = 3.65 MeV). 
The solar $^8$B neutrino background is proportional to the volume and can not be rejected by event identification, 
thus a small active volume is desirable. 
Because of the poor energy resolution, the tail of high energy $2\nu\beta\beta$ decay extends to the region of interest 
of $0\nu\beta\beta$ search. 
Therefore high light yield, high transparency of LS is required to improve the energy resolution.
%On the other hand, this type detector could contain huge amount of $\beta\beta$. 

Figure~\ref{fig:alowedregion} shows the allowed region of $0\nu\beta\beta$ decay calculated by neutrino oscillation parameters. 
Current ongoing projects search for ``degenerated mass'' region with $O(10^{1\rm - 2})$ kg $\beta\beta$ nuclei. 
In order to reach ``inverted hierarchy'' region including inverted mass hierarchy ($\nu_2 > \nu_1 > \nu_3$), degenerated and normal hierarchy ($\nu_3 > \nu_2 > \nu_1$), $O(10^3)$ kg $\beta\beta$ nuclei is needed. 
Liquid scintillator detectors which can contain large numbers of $\beta\beta$ nuclei are therefore one of the most sensitive.

\section{SNO+}

The SNO+ experiment plans to use tellurium loaded liquid scintillator (Te-LS) with hardware based on 
the SNO experiment. 
The liquid scintillator consists of Linear-Alkyl-Benzene and PPO(2g/L). 
780 tons of liquid scintillator in a 6 m radius acrylic vessel contains 0.5\% $^{\rm nat}$Te by weight.
The natural abundance of $^{130}$Te in $^{\rm nat}$Te is 34.1\%, thus this experiment does not use enriched material  
and 1,330 kg of $^{130}$Te can be loaded into the detector. 
The Q-value of $^{130}$Te is 2.527 MeV and this energy overlaps with environmental radiation from uranium 
and thorium as described above. 
While the acrylic vessel is not clean enough for the measurement, radiation from the vessel is rejected 
by using fiducialisation and self-shielding of Te-LS. 
From the current estimation, a 3.3 m radius fiducial volume (2.7 m self-shielding) makes the most sensitive condition. 
The detector position is $\sim$ 2,000 m below the ground level and it corresponds to 6,000 m water equivalent (m.w.e.). 
Due to the depth, cosmogenic muons come to the detector at the rate of only 70 events per day, and backgrounds from 
muon spallation products are negligible. 
Figure 2 shows expected background components and the energy spectrum. 
The main component of the background is the solar $^8$B neutrinos that are unavoidable and proportional to volume. 
The sensitivity is 1.9$\times$10$^{\rm 26}$ years at 90\% C.L. with 5 years of operations.
\begin{figure}[htbp]
 \begin{minipage}{0.5\hsize}
  \begin{center}
   \includegraphics[width=50mm]{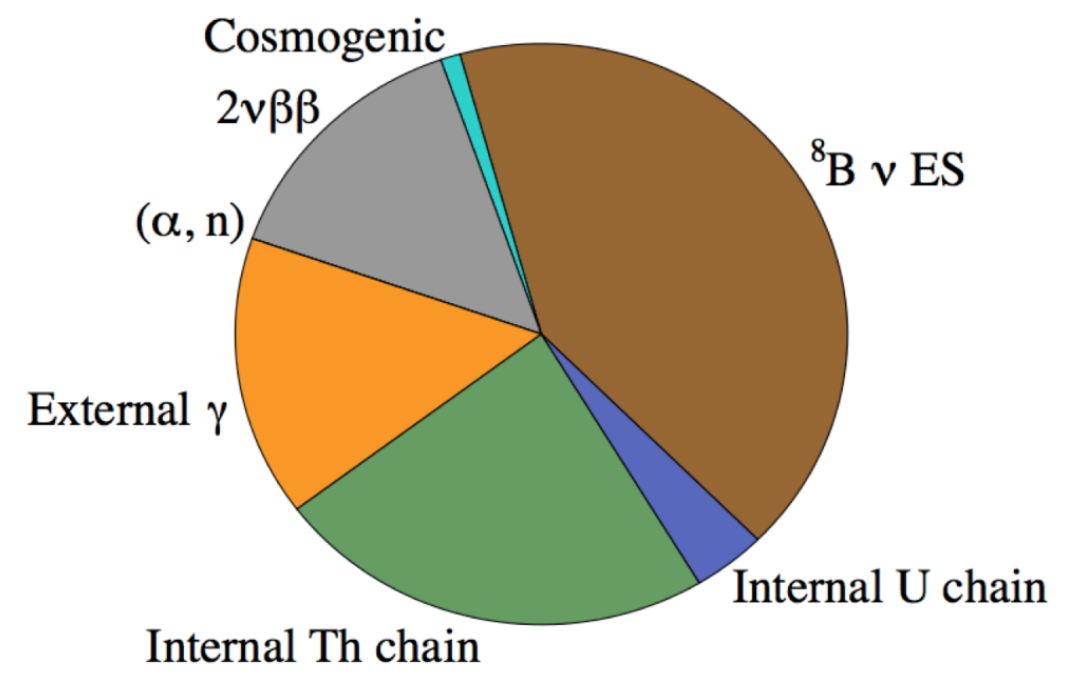}
  \end{center}
\label{fig:SNOspec}
 \end{minipage}
\begin{minipage}{0.5\hsize}
  \begin{center}
   \includegraphics[width=50mm]{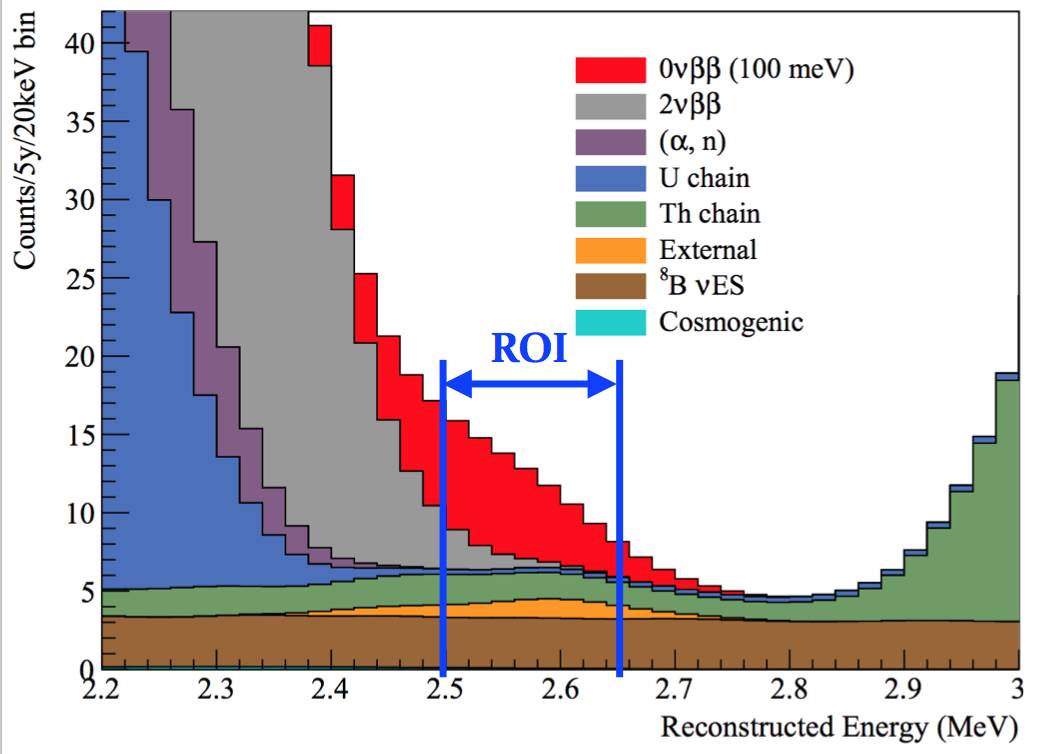}
  \end{center}
 \end{minipage}
 \vspace{-5mm}
   \caption{The left figure shows estimated background components of SNO+. The right plot shows expected energy spectrum of SNO+ with 5 years  of operation, 0.5\% $\rm ^{nat}$Te and R = 3.3 m fiducial volume~\cite{SNO+Nu2018}.}
 \end{figure}
The current stage of SNO+ is construction and test operations. 
There are three stages of the SNO+ $0\nu\beta\beta$ decay search: water phase, liquid scintillator phase, 
and tellurium loaded liquid scintillator phase. 
In order to investigate the external backgrounds, pure water was filled in the acrylic vessel and data acquisition was operated. 
The background level met their targets for future physics~\cite{MarkChen}.
SNO+ already terminated the water phase and released the results of solar $^8$B neutrino measurements~\cite{SNO+solar}, 
and of the search for invisible modes of nucleon decay~\cite{SNO+Ndecay}. 
Distilled LS filling of the acrylic vessel for liquid scintillator phase started in October of 2018. 
After the LS filling and several months of operations, tellurium will be loaded in the liquid scintillator. 
Tellurium can be dissolved in LS in the form of a Te-butanediol complex. 
3.8 tons of telluric acid was stored underground for more than 3 years 
in order to wait for the decay of long lived radioactivity made by cosmic ray muons and protons.
Construction of a tellurium acid purification plant is underway underground. 
SNO+ future plan called phase II includes the following improvements: 
1\% tellurium loading, high light yield and high transparency LS, 
detector upgrade with high quantum efficiency PMTs, concentrators replacement, inner bag,  
and $^{\rm 130}$Te enrichment.
SNO+ plans to cover the inverted hierarchy region.

\section{KamLAND-Zen}

\begin{figure}[htbp]
 \begin{minipage}{0.45\hsize}
  \begin{center}
   \includegraphics[width=35mm]{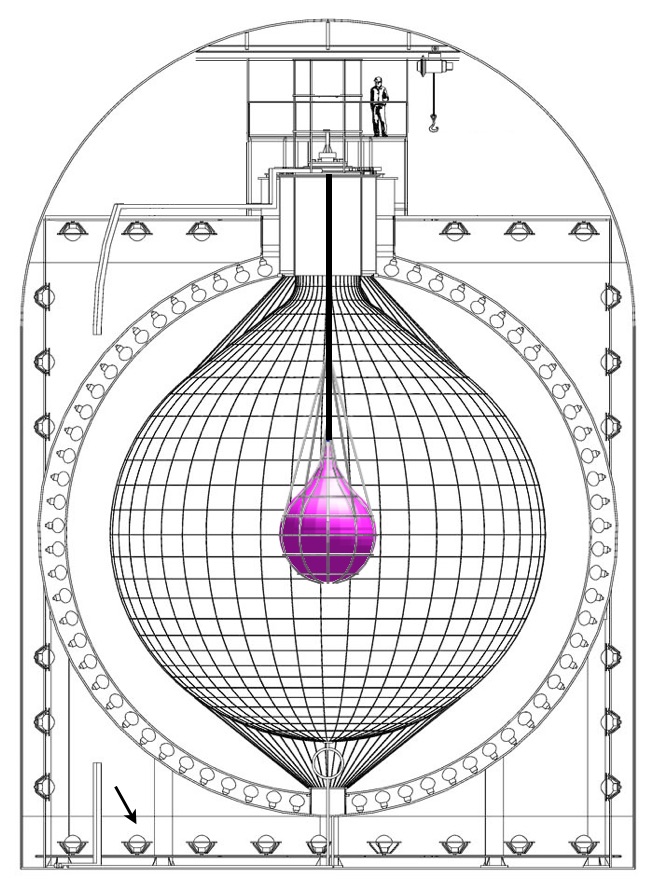}
  \end{center}
\vspace{-6mm}
  \caption{Schematic view of KamLAND-Zen}
  \label{fig:OneFig}
 \end{minipage}
\hspace{5mm}
 \begin{minipage}{0.45\hsize}
  \begin{center}
   \includegraphics[width=42mm]{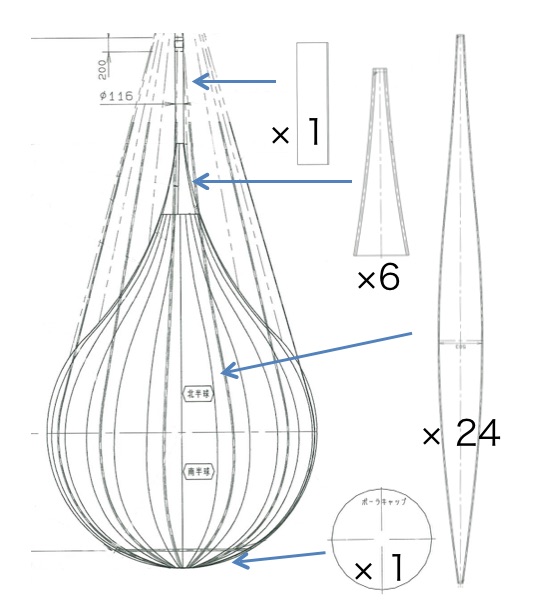}
  \end{center}
  \vspace{-6mm}
  \caption{Nylon film parts used in the mini-balloon}
  \label{fig:TwoFig}
 \end{minipage}
\end{figure}

KamLAND-Zen is a $0\nu\beta\beta$ decay experiment using $^{136}\rm Xe$ loaded liquid scintillator in the KamLAND detector~\cite{KL}. 
KamLAND is located is at 1,000 m depth (2,700 m.w.e.) where the cosmogenic muon rate is $\sim$0.3 Hz. 
In order to restrict the muon spallation products and solar $^8$B neutrinos backgrounds which are proportional to the volume, 
xenon loaded liquid scintillator (Xe-LS) is located in a nylon mini-balloon surrounded by 1,000 tons of LS contained 
in a 13 m diameter outer balloon (see Figure \ref{fig:OneFig}). 
Xe-LS can contain xenon at almost 3\% by weight and isotopic abundance of $^{136}\rm Xe$ is enriched to 90.6\%.
%The vertex and energy resolutions are 14.1 cm and 6.6 -- 7.3\%/$\sqrt{\rm E(MeV)}$ at $\sigma$ respectively.
KamLAND has achieved a 10$^{-17\sim -18}$ g/g contamination level for $^{238}\rm U$ and  $^{232}\rm Th$ in liquid scintillator\cite{solar}, thus its very clean container allows a high sensitivity search for $0\nu\beta\beta$ decay. 
The ``mini-balloon'' container for Xe-LS is made of 25 $\mu$m thickness nylon film 
and is suspended at the center of KamLAND as shown in Figure \ref{fig:TwoFig}. 
The nylon film has 99\% transparency and a contamination level of $\sim$2$\times$10$^{-12}$ g/g 
level for $^{238}\rm U$ and  $^{232}\rm Th$ .
To make the drop-shaped container, nylon films were cut for each part and connected by heat welding 
in a class-1 super clean room.

\subsection{KamLAND-Zen 400}
\begin{figure}[htbp]
 \begin{minipage}{0.45\hsize}
  \begin{center}
   \includegraphics[width=55mm]{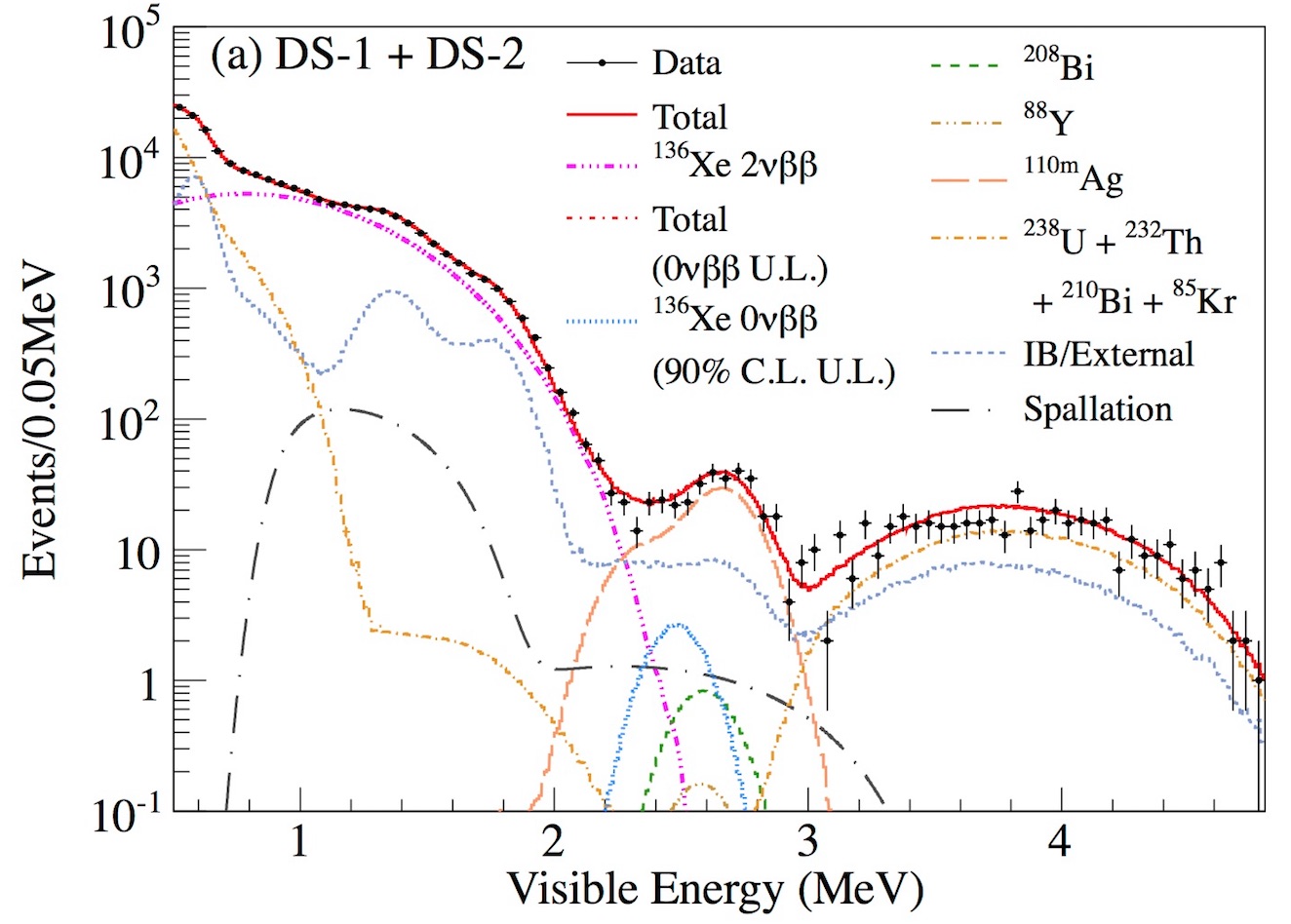}
  \end{center}
\vspace{-8mm}
  \caption{Energy spectrum in Phase-I (R $<$ 1.35 m)}
  \label{fig:ThreeFig}
 \end{minipage}
 \hspace{5mm}
 \begin{minipage}{0.45\hsize}
  \begin{center}
   \includegraphics[width=55mm]{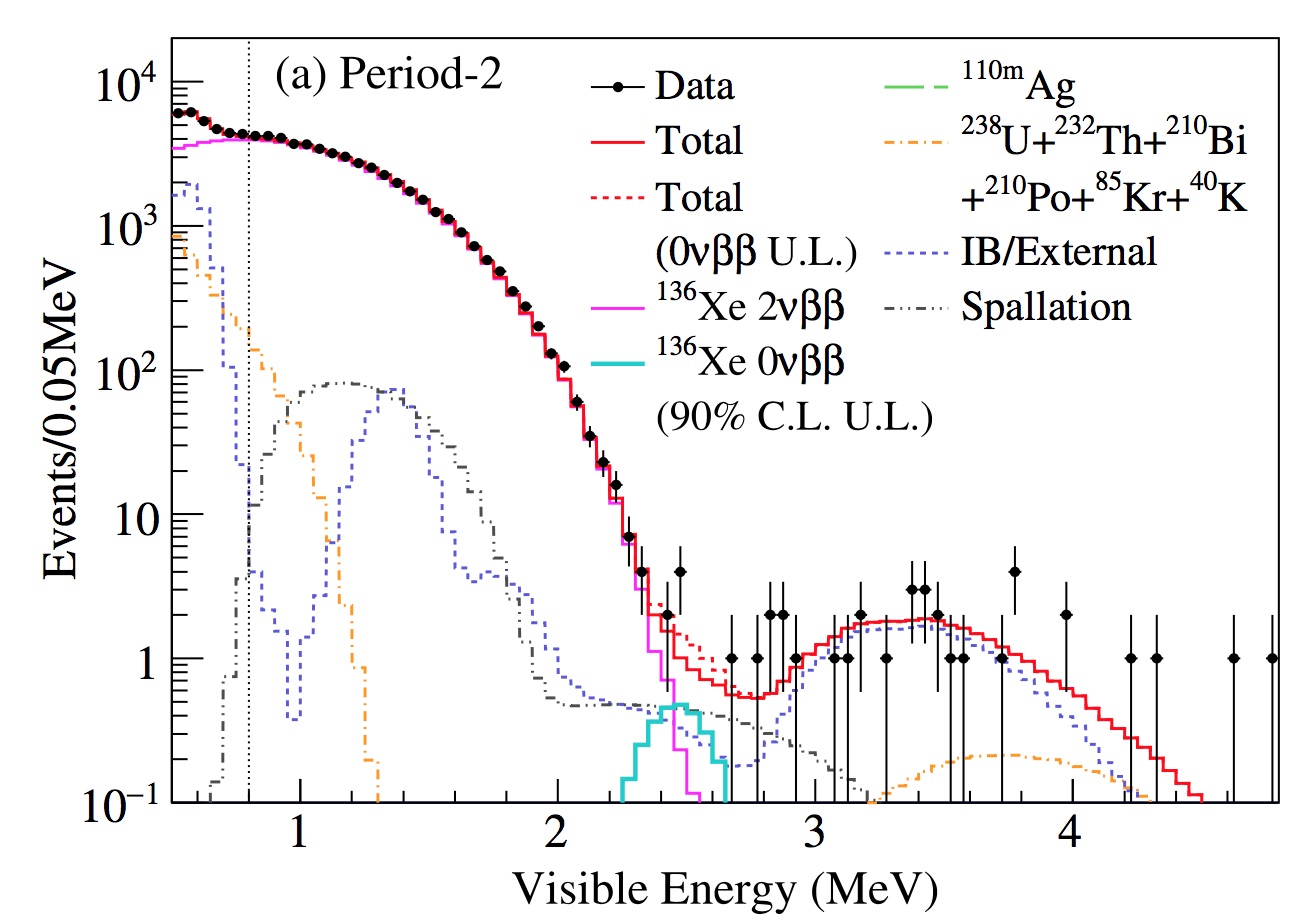}
  \end{center}
\vspace{-8mm}
  \caption{Energy spectrum in latter period of Phase-II (R $<$ 1.0 m)}
  \label{fig:FourFig}
 \end{minipage}
\end{figure}

KamLAND-Zen 400 started data acquisition in October 2011 and terminated in October 2015, 
including a purification period from June 2012 to December 2013. 
In the first phase before purification (Phase-I), we found $^{110m}\rm Ag$ events 
in the region of interest for $0\nu\beta\beta$ decay (see Figure \ref{fig:ThreeFig})\cite{KLZen1st}.
We suspect the impurities came from fall out of the Fukushima reactor accident.
After Phase-I, we extracted xenon from LS and purified it by distillation and getter filtering.
After the removal of xenon, LS was purified three times by distillation, and replaced by new one twice.
Unfortunately, the inner surface of the mini-balloon was contaminated from mine air by a pump failure,
thus limiting the effective fiducial volume in phase II. 
The energy spectrum of purified Xe-LS in Figure \ref{fig:FourFig} shows no $^{110m}\rm Ag$ peak.
From the combined analysis with Phase-I and Phase-II data, a lower limit for the half life of $0\nu\beta\beta$ decay is 
$\rm T_{1/2} >1.07\times 10^{26}$ years at 90\% C.L 
corresponding to $\langle m_{\beta \beta}  \rangle <$ 61$-$165 meV~\cite{Zen400final}.

\subsection{KamLAND-Zen 800}

Due to the $\gamma$-rays from surface contamination of the mini-balloon, the sensitivity was restricted in KamLAND-Zen 400.
Therefore we started KamLAND-Zen 800 project with almost 750 kg xenon and cleaner mini-balloon. 
In order to make a cleaner mini-balloon, we applied a number of techniques: clean wear control, particle flow check, 
static-electricity control by ion generation devices and humidity control, 
%film cover setting for mini-balloon film, 
using a film cover to protect the mini-balloon film, 
and the introduction of a semi-automatic welding machine.
We used three clean wear layers: a clean inner suit, the first clean suit wearing in a class-1,000 clean room, 
and second clean suit changing in a class-1 super clean room. 
The mini-balloon was constructed in a separate super-clean room.
Custom order nylon film is easily charged, and the static-electricity collects dusts including environmental radioactivities.
Static-electricity is prevented by 65\% humidity in general, therefore a mist generation system was set just before the ULPA filter.
We also applied protective nylon film covers for the mini-balloon nylon film.
When we welded films and did leak check for welding lines, the mini-balloon films were protected from 
dust contamination by the cover films. 
For the KamLAND-Zen 400 mini-balloon, welding was done by a hand pressing machine. 
When we used this machine, a person had to keep their body on the nylon film, 
%and there were possibility to drop dusts from wear or human.
and dust from the clean suit or the person could drop on the film.
Thus we introduced a semi-automatic welding machine 
to avoid dust drop and press weight differences by each person.

The mini-balloon production was done by the following procedure: 
nylon film was washed by ultra pure water with ultra sonic cleaning to reject initial surface contamination, 
film cover setting for mini-balloon film, clipping to each part, film connection by welding, 
leak check by helium gas and helium detector, and repairing holes by glue.

\begin{figure}[htbp]
 \begin{minipage}{0.5\hsize}
  \begin{center}
   \includegraphics[width=35mm]{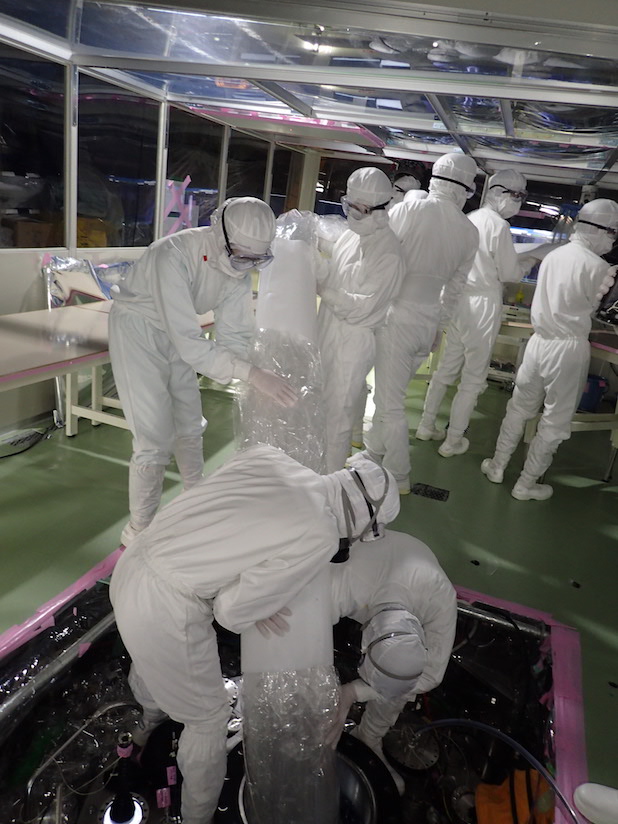}
  \end{center}
  \caption{mini-balloon installation}
  \label{fig:FiveFig}
 \end{minipage}
 \begin{minipage}{0.5\hsize}
  \begin{center}
   \includegraphics[width=70mm]{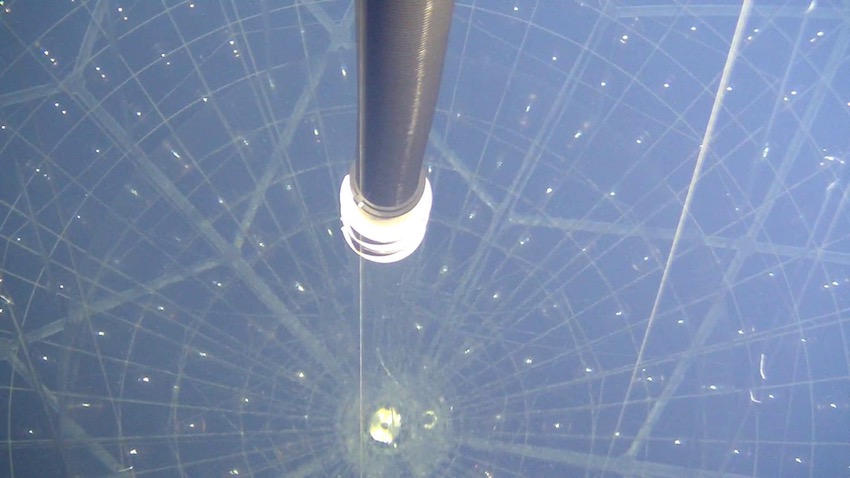}
  \end{center}
  \caption{Expanded mini-balloon in KamLAND}
  \label{fig:SixFig}
 \end{minipage}
\end{figure}

Installation of the mini-balloon to KamLAND was done on May 10, 2018 (see Figure \ref{fig:FiveFig}).
For the install preparations in Kamioka site, we set up a class-50 level clean room at the top of the KamLAND detector.
Due to the spherical shape of the detector and access point to inside the outer balloon being only 50 cm in diameter,  
we folded the mini-balloon keeping the shape using perforated teflon sheet and teflon tubes.
We applied cover nylon films between the mini-balloon film and teflon sheets to avoid damage during the installation process.
We installed the mini-balloon with heavier LS (+0.4\%) compared to the KamLAND LS density.
After sinking of mini-balloon in KamLAND LS, teflon sheets and cover nylon films were removed and pulled up.
After the installation, we filled slightly heavier LS (+0.015\%) without xenon, and the mini-balloon was expanded as shown in Figure \ref{fig:SixFig}.

We purified non-xenon-loaded LS in the mini-balloon by distillation after the installation because $^{232}\rm Th$ level 
was slightly high O(10$^{-15}$) g/g. 
After the purification, xenon dissolving to LS was performed from December 2018 to January 2019. 
KamLAND-Zen 800 data acquisition was started in January 2019. 
It is expected that after 3 years of data taking the inverted hierarchy region will be probed. 

\subsection{KamLAND2-Zen}

Possible backgrounds for the future project of KamLAND2-Zen are $^{\rm 10}$C, $^{\rm 214}$Bi, 
solar $^8$B neutrinos, and $2\nu\beta\beta$.
To reject these backgrounds, we have the following efforts: electronics upgrade for $^{\rm 10}$C rejection, 
scintillation balloon development~\cite{ScintiFilm} for $^{\rm 214}$Bi rejection, 
and particle ID by imaging devices for $^{\rm 10}$C and  $^{\rm 214}$Bi rejection. 
Mini-balloon is located at almost 10 m depth in KamLAND, thus it has 1.8 atmosphere pressure. 
It means that more xenon can be dissolved by Henry's law. 
We are considering improvement of the xenon/LS ratio in the mini-balloon to restrict $^{\rm 10}$C 
and solar $^8$B neutrinos backgrounds which are proportional to volume. 
$2\nu\beta\beta$ background could be restricted by energy resolution improvement. 
In order to accomplish these goals, we have studied high light yield LS, high Q.E. PMTs with light collection mirrors.
Currently some development activities are on going and we are preparing for budget requests. 

\section{Summary}

Liquid scintillator based experiments have good sensitivities to search for Majorana neutrino mass 
in the inverted hierarchy region of $0\nu\beta\beta$ decay based on ultra low background environments. 
SNO+ experiment is ongoing with LS filling of the detector. 
After the filling and several month operations, tellurium will be introduced in LS and data acquisition will be started. 
The lower limit for the half life of $0\nu\beta\beta$ of $^{136}\rm Xe$ is $\rm T_{1/2} >1.07\times 10^{26}$ years 
at 90\% C.L. using the KamLAND-Zen 400 experiment,  
corresponding to $\langle m_{\beta \beta}  \rangle <$ 61$-$165 meV.
In the next phase, KamLAND-Zen 800 was started with significant improvements. 
Preparation for KamLAND2-Zen has begun. 

\Acknowledgements
I am grateful to Prof. Mark Chen who gave me the figures and informations about SNO+ experiment.

\end{document}